\documentclass[12pt]{article}
\usepackage{amssymb}
\usepackage[dvips]{graphicx}
\usepackage{amsmath}

\begin{document}

\newcommand{\be}{\begin{eqnarray}}
\newcommand{\ee}{\end{eqnarray}}
\renewcommand{\ni}{\noindent}
\newcommand{\bdm}{\begin{displaymath}}
\newcommand{\edm}{\end{displaymath}}
\newcommand{\gf}[1]{f^{[#1]}}
\newcommand{\igf}[1]{g^{[#1]}}
\newcommand{\azl}[1]{AZ_{#1}}

\newtheorem{theo}{Theorem}
\newtheorem{cor}[theo]{Corollary}
\newtheorem{defn}{Definition}

\title{Two Dimensional Directed Lattice Walks with Boundaries}
\author{Arvind Ayyer\\
\small Department of Physics\\
\small 136 Frelinghuysen Rd\\
\small Piscataway, NJ 08854.\\
\small \texttt{ayyer@physics.rutgers.edu}\\
\\
Doron Zeilberger\\
\small Department of Mathematics\\
\small 110 Frelinghuysen Rd\\
\small Piscataway, NJ 08854.\\
\small \texttt{zeilberg@math.rutgers.edu}\\
}
\date{}

\maketitle

\begin{abstract}
We present general algorithms (fully implemented in Maple) for calculations of various quantities related to constrained directed walks for a \emph{general} set of steps on the square lattice in two dimensions. As a special case, we rederive results of earlier works.
\end{abstract}

\section{Introduction}
Lattice walks form some of the most well-studies problems in combinatorics. Studies of walks with simple set of steps such as $\{ (1,1), (1,-1) \}$ (Dyck or Catalan steps), $\{ (1,1),(1,0),(1,-1) \}$ (Motzkin steps) and $\{ (1,1), (1,-1),\\ (2,0) \}$ (Schroeder steps) reveal connections with many classical combinatorial problems.

On the other hand, one can hope to find combinatorial problems of all kinds within lattice walks. In particular, there are walks whose enumerating generating functions are rational; there are some which are algebraic; there are some which are holonomic; and finally, some are not even that - something more general? Who knows?

We embark on a program of automating the study of lattice walks with the \emph{small-step-but-giant-leap} of automating two dimensional directed lattice walks with boundaries. These are walks which are bounded in one direction and which take each step from a finite set $S$ such that every step in $S$ has a strictly positive inner product with the unbounded direction.

The motivation for this study comes from the statistical mechanical study of polymers held between two close plates \cite{R}. Consider a long linear polymer in dilute solution constrained between two plates. Naively one would expect the polymer to exert a force on the plate, simply because the polymer would not naturally prefer to be confined. However, this can change if there is an interaction between the plates and the polymer. In the latter case, as we tune the strength of this interaction, a phase transition can occur which would change the sign of the overall interaction. 

This is a hard problem to tackle, because one would have to calculate quantities of physical or mathematical interest separately for each model one considers. However, a nice framework for modelling the polymer is through a directed walk on a lattice. This is self-avoiding by definition and can therefore represent the rough configuration of the polymer even though the microscopic details may differ.

So far this framework has been used to study walks with a specific set of steps given by $(1,1),(1,-1)$ \cite{BORW}. In general, this is not quite satisfactory for a couple of reasons. Firstly, this set of steps allows the polymer to move at an angle of $\pm 45^\circ$ only and thus severely restricts its configuration space. For example, a better model would be $(1,2),(2,1),(1,-2),(2,-1)$.

Secondly, one would like to consider polymers with different bond lengths. In other words, suppose that the polymer has various molecules $\{ X_i \}$. And the values of the bond lengths between the various species is $\{ l_{ij} = |X_i-X_j| \}$, where $|X_i - X_j|$ denotes the bond length between $X_i$ and $X_j$.

The simplest way to incorporate parameters representing the interaction with the walls in these models is to write each walk as a monomial in two parameters $t,s$ where the power of $t$ is precisely the number of times the walk has touched the bottom wall and similarly, the power of $s$ is for the top wall.

Until now, the study of the constrained walk with an arbitrary set of bond lengths has been daunting because there are no general results in this direction. We present a toolbox of algorithms (implemented in a Maple package named \emph{POLYMER}) which can be used as a black box to calculate various quantities of interest such as the generating function and the free energy as well as plot quantities such as the force in different regimes. We emphasize that these calculations can be done for \emph{any} choice of steps at least in principle. In practice, of course, one is restricted by the limited resources of the largest computers.

We should mention that, unknown to us, the equations for infinite width walks were already written down in the elegant piece of work, albeit in slightly different language, in \cite{Du}. While this work was in preparation, there also appeared \cite{Bo} where the topic of constrained walks with arbitrary steps is also treated. The flavor of the work is somewhat different there. While the results there are of much greater generality, they are existence results. The results in this paper are more algorithmic in nature.

\section{The Setting}
\begin{figure}[h!]
\begin{center}
\includegraphics[width=10cm]{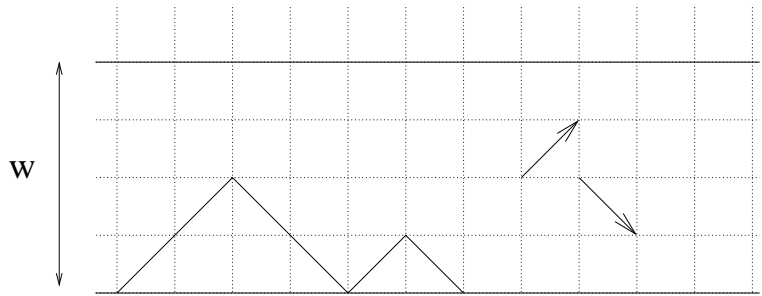}
\caption{A walk with width $w=4$ involving the steps $(1,1)$ and $(1,-1)$}
\label{latdyck}
\end{center}
\end{figure}

Consider a walk constrained in the two dimensional square lattice $\mathbb{Z}^2$by $0 \leq y \leq w$ where $w$ is the width. In Fig. \ref{latdyck} we show a walk with width $w=4$ involving the steps $(1,1)$ and $(1,-1)$.

Such a walk can always be reinterpreted as a walk in the region given by $0 \leq x-y \leq w,y \geq 0$. This is done simply by rotating and reflecting the above figure in the appropriate way. The line $x-y=0$ is the same as $y=0$ above and the line $x-y=w$ becomes $y=w$. 

\begin{figure}[h!]
\begin{center}
\includegraphics[width=10cm]{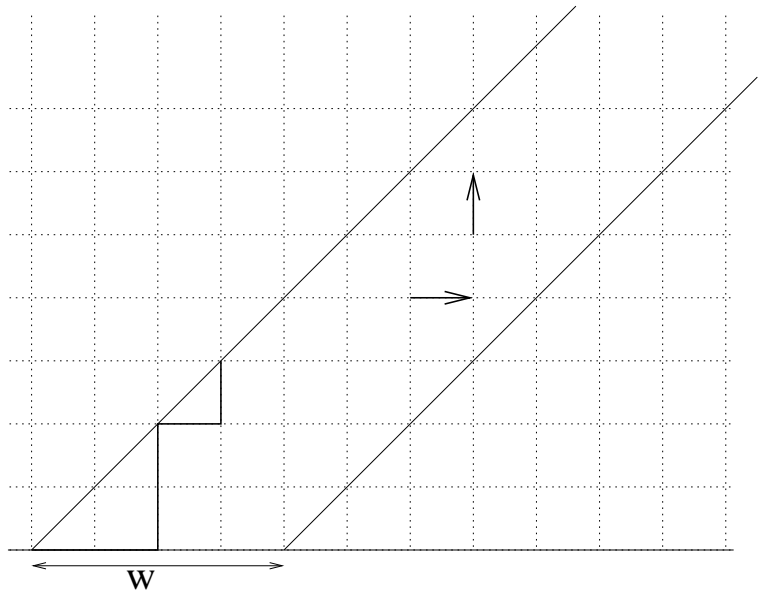}
\caption{The same ballot walk with $w=4$ with steps $[1,0]$ and $[0,1]$}
\label{baldyck}
\end{center}
\end{figure}

In Fig. \ref{baldyck} the same walk is redrawn for this situation. Notice that the steps have also been rotated. They have become $[1,0]$ and $[0,1]$. The former walks are known as \emph{Dyck paths} and because the latter represent the so-called ballot problem, we call them \emph{ballot paths}. For convenience, we represent Dyck steps with the usual parentheses $(,)$ and ballot steps with square parentheses $[,]$. Points on the lattice are always denoted by the usual parentheses $(,)$.

Since we are interested in \emph{directed} walks, the steps should have a positive inner product with the vector $(1,0)$ for Dyck paths and the vector $[1,1]$ for ballot paths.


\section{Calculating Walks}
The main idea in counting the number of walks with a general set of steps is \emph{recursion}.

\subsection{Simple Walks}
Consider the number of ballot paths from $(0,0)$ to the point $(m,n)$ which we denote as $c(m,n)$. Let us first consider simple steps. For example, $[1,0]$ and $[0,1]$. Then $c(m,n)$ satisfies the simple recurrence relation
\be
c(m,n) & = & 0  \quad \textrm{if} \; m<0,n<0,m-n<0 \; \textrm{or} \; m-n>w, \cr
c(m,n) & = & c(m,n-1) + c(m-1,n) \quad \textrm{if} \; m>n>0, 
\ee
\noindent
because using the given steps, one can arrive at the point $(m,n)$ either by the step $[1,0]$ from $(m-1,n)$ or by the $[0,1]$ steps from $(m,n-1)$ if $m>n>0$ and hence the number of such walks simply adds. On the other hand, one can never reach $(m,n)$ if $m<0,n<0,m-n<0$ or $m-n>w$. Thus, we only need the initial condition $c(0,0)=1$ (the null walk) to determine all walks. For example,
\be
c(2,1) & = & c(2,0) + c(1,1) \cr
& = & (c(1,0)+c(2,-1)) + (c(1,0)+c(0,1)) \cr
& = & (c(0,0)+0) + (c(0,0)+0) \cr
& = & 2
\ee
\noindent
and one can easily check that there are only two walks from $(0,0)$ to $(2,1)$.
\noindent
Let us take a more complicated example. Suppose the steps are $[1,0],[0,2]$ and $[1,1]$. Then the recurrence becomes more complicated and is given by
\be
c(m,n) & = & 0  \quad \textrm{if} \; m<0,n<0,m-n<0 \; \textrm{or} \; m-n>w, \cr
c(m,n) & = & c(m,n-2)+c(m-1,n)+c(m-1,n-1) \quad \textrm{if} \; m>n. 
\ee

When there is yet another constraint given by $x-y \leq w$, we simply need to put in another ``if'' condition and the main recurrence relation remains unchanged.

This idea can be implemented as an algorithm as follows: Let $Steps$ represent the set of all possible steps and let $Steps[i]$ denote the $i$th step. Then the number of walks $c(m,n)$ from $(0,0)$ to $(m,n)$ with width $w$ is calculated as follows:

\begin{center}

\ni if $m=n=0$ then\\
\indent RETURN 1

\ni if $m+n<0$ then \\
\indent RETURN 0

\ni if $m-n<0$ or $m-n>w$ then \\
\indent RETURN 0

\ni if $m>n>0$ and $m-n<w$ then \\
\indent $\quad Prev = \{(m,n)-Steps[i]| i=1,2,...\}$ \\
\indent $\displaystyle \quad RETURN \sum_i c(Prev[i])$ 

\end{center}

\subsection{Walks with Boundary Interactions}
In combinatorics, boundary interactions can be implemented with variables $t,s$ which ``record'' the number of times the walk touches the left and right walls respectively. More precisely, each walk is assigned a monomial in $t$ and $s$ where the degree of the monomial in $s(t)$ is the number of times the walk touches the left wall (right wall). The initial point does not count.

For example, consider walks from $(0,0)$ to $(2,2)$ with steps $[1,0],[0,1]$. There are only two possibilities as shown.


\begin{figure}[h!]
\centering
\begin{tabular}{cc}
\hspace{-1cm}
\includegraphics[width=5cm]{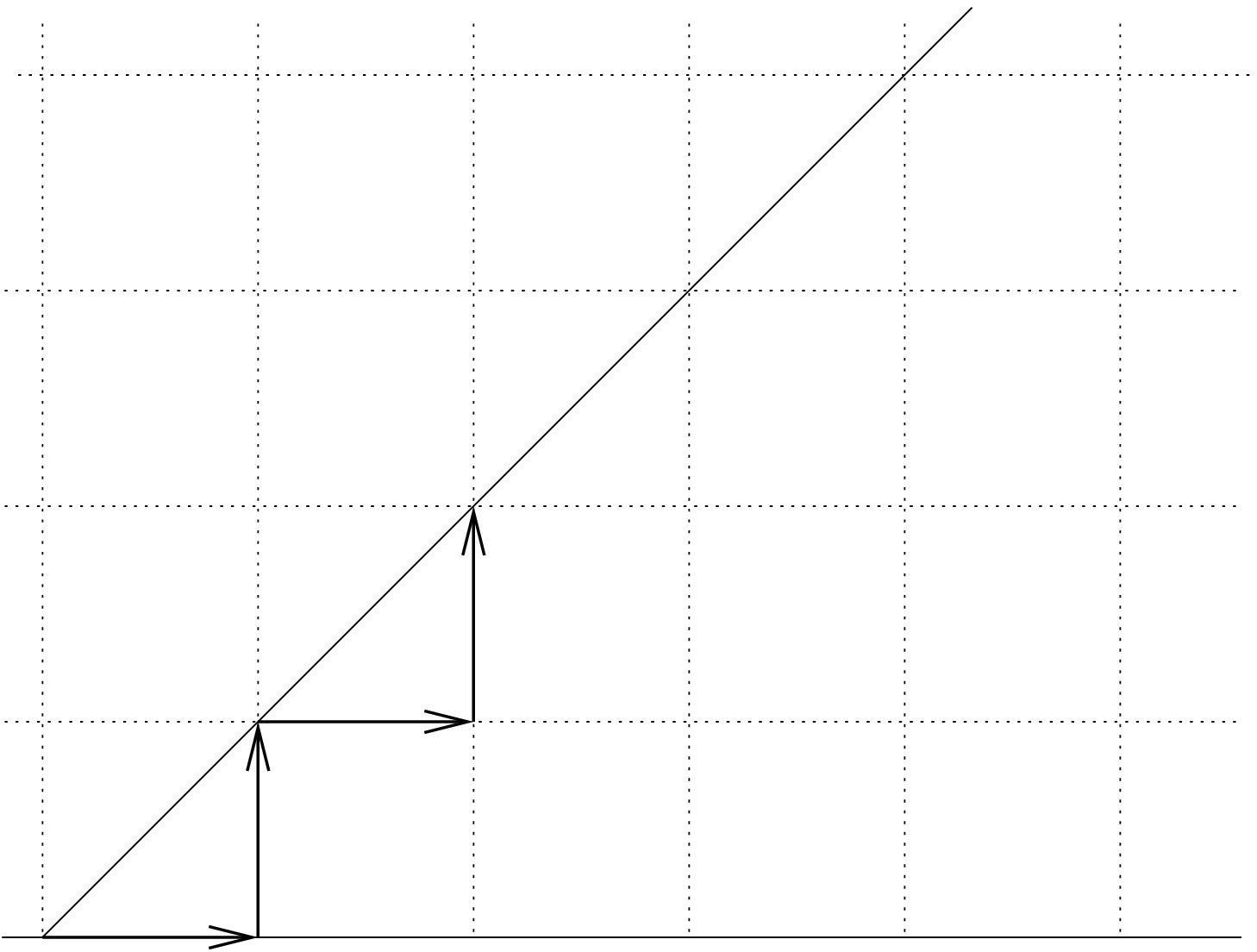} &
\includegraphics[width=5cm]{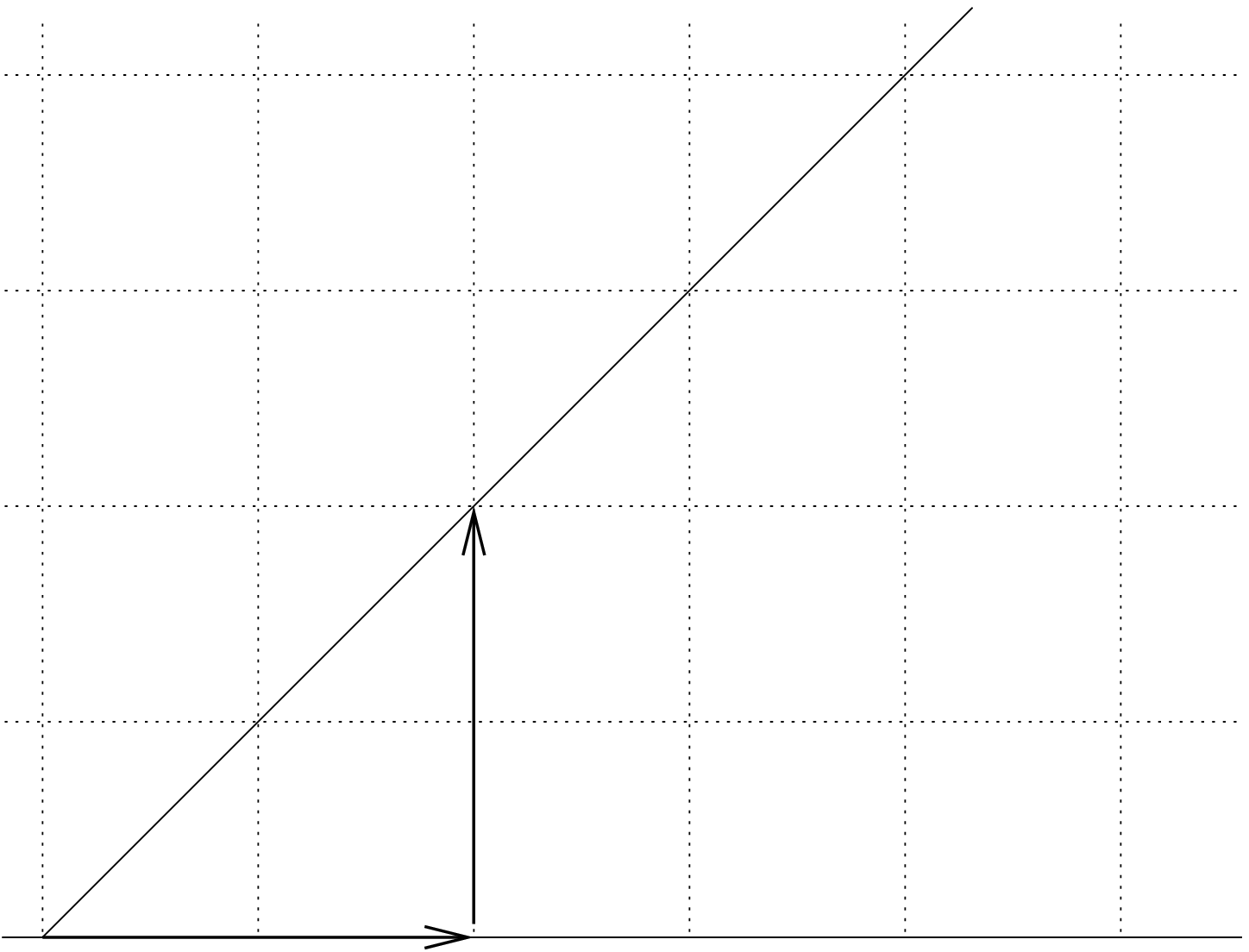}
\end{tabular}
\caption{The only possible walks from $(0,0)$ to $(2,2)$ using the steps $[0,1]$ and $[1,0]$}
\label{walk0to2}
\end{figure}

When $w=1$, only the left walk is permitted and carries a weight of $t^2 s^2$ because it touches both walls twice. When $w=2$, both walks are permitted but carry different weights. The left one carries a weight of $t^2$ and the right one, a weight of $ts$. When $w=3$, the left one still carries a weight of $t^2$ but the right one does not touch the right wall at all and therefore carries a weight of $t$. Thus,
\be
c_1(t,s) & = & t^2 s^2 \\
c_2(t,s) & = & t^2 + ts \\
c_3(t,s) & = & t^2 + t
\ee

Setting $t=s=1$ gives the number of such walks. The algorithm representing these walks closely resembles simple walks. 
\begin{center}
\ni if $m=n=0$ then\\
\indent RETURN 1

\ni if $m+n<0$ then \\
\indent RETURN 0

\ni if $m-n<0$ or $m-n>w$ then \\
\indent RETURN 0

\ni if $m-n=0$ then \\
\indent $\quad Prev = \{(m,n)-Steps[i]| i=1,2,...\}$ \\
\indent $\displaystyle \quad RETURN \sum_i(t \cdot c(Prev[i]))$ 

\ni if $m-n=w$ then \\
\indent $\quad Prev = \{(m,n)-Steps[i]| i=1,2,...\}$ \\
\indent $\displaystyle \quad RETURN \sum_i(s \cdot c(Prev[i]))$ 

\ni if $m>n>0$ and $m-n<w$ then \\
\indent $\quad Prev = \{(m,n)-Steps[i]| i=1,2,...\}$ \\
\indent $\displaystyle \quad RETURN \sum_i(c(Prev[i]))$ 

\end{center}

\section{Generating Functions}
The generating function is an important tool in combinatorics. It is another way to package the same information as brute-force counting. It is a formal power series whose coefficients in the taylor series give precisely the count. More specifically, if $c_w(n)$ is the number of ballot walks from $(0,0)$ to $(n,n)$ with width $w$, the generating function $\phi_w(z)$ is given by
\be
\phi_w(z) = \sum_{n=0}^\infty c_w(n) z^n
\ee

We demonstrate the calculation of the generating function for both the case of finite width and infinite width. The ideas involved in both are quite different and so they will be treated differently. However, both involve a set of tricks commonly used in combinatorics. We describe them in detail in subsequent sections.




\subsection{Finite width}
We use the same ideas described above to calculate the generating function of the number of walks with any set of steps.

We first start with a nontrivial example. We will spell out all the details here. Consider the steps $[1,0],[0,2],[1,1]$ and width, $w=2$. We define three generating functions $\phi_i(z),i=0,1,2$ for this problem, where $\phi_i(z)$ counts the number of walks from $(0,0)$ to $(n+i,n)$.

In terms of initial conditions, only $\phi_0$ has a nontrivial condition, namely the zero-step walk from $(0,0)$ to $(0,0)$. Now, let us consider each of the generating functions one at a time.

For a walk ending at $(n,n)$, the last step can have two possibilities. It can either end with a $[1,1]$ step from $(n-1,n-1)$ (which is also described in terms of $\phi_0$) or it can end with a $[0,2]$ step from $(n,n-2)$ (which is described by $\phi_2$). Since the weight of the walk depends only on the $y$-coordinate, we have the following equation for $\phi_0$
\be
\phi_0 = 1 + z \phi_0 + z^2 \phi_2
\ee

Now, for a walk ending at $(n+1,n)$, there are two ending possibilities. The final step can be $[1,0]$ from $(n,n)$ (described by $\phi_0$) or it can be $[1,1]$ from $(n,n-1)$ (described by $\phi_1$). The $[0,2]$ step is not possible because the walk would have to start outside the prescribed strip. Thus,
\be
\phi_1 = z \phi_1 + \phi_0
\ee

Finally, a walk ending at $(n+2,n)$ also has two possibilities. The final step can be $[1,0]$ from $(n+1,n)$ (described by $\phi_1$) or it can be $[1,1]$ from $(n+1,n-1)$ (described by $\phi_2$). 
\be
\phi_2 = \phi_1 + z \phi_2
\ee

These are now three linear equations in three variables. These, when solved, for $\phi_0$ gives the rational function
\be
\phi_0 = \frac{1-2z+z^2}{1-3z+2z^2-z^3}
\ee

This example contains the essence of the argument to follow. For any set of steps and width $w$, we will always have $w+1$ linear equations in the variables $\phi_0,\cdots,\phi_w$ independent of the number and kinds of steps involved. These equations will be linear precisely because the ending of each walk contributing to $\phi_i$ takes its last step starting from some other walk contributing to, say, $\phi_j$. To be precise, for $i=0,\cdots,w$,
\be
\phi_i & = & \delta_{i,0}+ \sum_{\substack{j \\ 0 \leq i+Steps[j]_y-Steps[j]_x \leq w}} z^{Steps[j]_y} \phi_{i+Steps[j]_y-Steps[j]_x}
\ee

Solving this system will always lead to \emph{rational} solutions for each of the generating functions $\phi_i$. And modulo unexpected cancellations, they all have the same denominator - the determinant of the corresponding matrix in the linear system!


\subsection{Finite Width with Boundary Interactions}
We can also calculate the generating function of walks with variables $t,s$ (called \emph{weight enumerators}) using essentially the same idea.

Let us consider the same example with steps $[1,0],[0,2],[1,1]$ and width, $w=2$. We again have three generating functions $\phi_i(z),i=0,1,2$. Since we have a factor of $t$ everytime we touch the left wall, the equation for $\phi_0$ is 
\be
\phi_0 = 1 + t z \phi_0 + t z^2 \phi_2
\ee

The equation for $\phi_1$ is unchanged because walks contributing to it are touching neither of the two walls at the last step.
\be
\phi_1 = z \phi_1 + \phi_0
\ee

The equation for $\phi_2$ is affected because walks contributing to it are exactly on the right wall at the final step.
\be
\phi_2 = s \phi_1 + s z \phi_2
\ee

Solving this gives
\be
\phi_0 = \frac{(1-z)(1-sz)}{1-z-sz-tz+sz^2+tz^2-stz^3}
\ee

As this example shows, the generating functions for these walks can also be calculated exactly using the same technique as in the previous section. In fact, the equations are quite similar unweighted enumeration. For the extreme generating functions, the modified equations look like
\be
\phi_0 & = & 1 + \sum_{\substack{j \\ 0 \leq Steps[j]_y-Steps[j]_x \leq w}} t z^{Steps[j]_y} \phi_{Steps[j]_y-Steps[j]_x} \\
\phi_w & = & \sum_{\substack{j \\ 0 \leq w+Steps[j]_y-Steps[j]_x \leq w}} s z^{Steps[j]_y} \phi_{w+Steps[j]_y-Steps[j]_x}
\ee
\noindent
while for the remainder, ie $i=1,\cdots,w-1$, the equations remain the same.
\be
\phi_i & = & \sum_{\substack{j \\ 0 \leq i+Steps[j]_y-Steps[j]_x \leq w}} z^{Steps[j]_y} \phi_{i+Steps[j]_y-Steps[j]_x}
\ee

\subsection{Infinite Width}
We have to manipulate generating functions in a different way to calculate them for the infinite width walks. We will need a number of definitions for this purpose.

First off, define an \emph{$[ij]$ walk} as one which starts on the line $x-y=i$ and ends on the line $x-y=j$. Since we have infinite width, both $i,j \geq 0$. Denote the generating function of such walks by $\gf{ij}(z)$. 

Define an \emph{irreducible $[ij]$ walk} as a walk which, as before, starts on the line $x-y=i$ and ends on the line $x-y=j$ with the restriction that it touches the line corresponding to the minimum of $i$ and $j$ only at the corresponding endpoint. Denote the generating function of such an irreducible walk by $\igf{ij}(z)$.

Now the idea is to relate these generating functions for different values of $i$ and $j$ where both range from 0 to a certain finite value depending on the kind of steps. 

Consider the following set of steps: $\{ [0,1],[1,0],[2,0],[0,2] \}$. First off, a $[00]$ walk is either the empty walk or it is composed of an irreducible $[00]$ walk  followed by a smaller $[00]$ walk.

\be
\gf{00} & = & 1 + \gf{00} \igf{00} \label{f00}
\ee

Next, a $[01]$ walk is always uniquely composed of an arbitrary $[00]$ walk followed by an irreducible $[01]$ walk. Similarly, a $[10]$ walk is uniquely composed of an irreducible $[10]$ walk followed by an arbitrary $[00]$ walk. 

\be
\gf{10} & = & \igf{10} \gf{00} \label{f01}\\
\gf{01} & = & \igf{01} \gf{00} \label{f10}
\ee

A $[11]$ walk either never goes below the first level, in which case it is simply the same as a $[00]$ walk, or if it does, it is composed of an irreducible $[10]$ walk followed by an arbitrary $[01]$ walk. 

\be
\gf{11} & = & \gf{00} + \igf{10} \gf{01} \label{f11}
\ee

Now, we go on to describe the irreducible walks. In each case, we have to consider different cases for the starting step and the ending step. First, an irreducible $[00]$ walk can begin with either the $[1,0]$ or $[2,0]$ step and end with either the $[0,1]$ or $[0,2]$ step. If the walk starts with $[1,0]$ and ends with $[0,1]$, then there could be an arbitrary $[00]$ walk in between. If the walk starts with $[1,0]$ and ends with $[0,2]$, there has to be an arbitrary $[01]$ walk in between. If the walk starts with $[2,0]$ and ends with $[0,1]$, there has to be an arbitrary $[10]$ walk in between. And finally, if the walk starts with $[2,0]$ and ends with $[0,2]$, there is a $[11]$ walk in between. For each of these cases only the $y$-coordinate of the steps give the corresponding powers of $z$.

\be
\igf{00} & = & z \gf{00} + z^2 \gf{01} + z \gf{10} + z^2 \gf{11} \label{g00}
\ee

For an irreducible $[01]$ walk, we just need to consider the starting steps. If it starts with $[1,0]$, the remainder is an arbitrary $[00]$ walk. If it starts with $[2,0]$, the remainder is again an arbitrary $[10]$ walk. A very similar argument on the ending step yields the equation for an irreducible $[10]$ walk.

\be
\igf{01} & = &  \gf{00} + \gf{10} \label{g01}\\
\igf{10} & = & z \gf{00} + z^2 \gf{01} \label{g10}
\ee

This finally gives the desired seven equations in seven variables. Notice that all the equations are algebraic and therefore the solution of this system will also be algebraic, ie, the solution of a polynomial equation. We are ultimately interested in $\gf{00}$ and in this case, eliminating the other variables and replacing $\gf{00}$ by $F$ gives
\be
z^4 F^4 - 2 z^3 F^3 - z^2 F^3 + 2 z^2 F^2 + 3 z F^2 - 2 z F + 1-F = 0
\ee

For the generic set of steps, we have the following algorithm. Let 
\be
m = max(|Steps[i]_x-Steps[i]_y|)
\ee

Then we define $\gf{ij}$ and $\igf{ij}$ for $i,j=0,1,\cdots,m-1$. It is these generating functions we will write down equations for. It turns out the equations for $\gf{ij}$ are almost completely independent of the kind of steps. In particular, only $\gf{00}$ depends on whether there are any steps of the form $[k,k]$.
\be
\gf{00} = 1 + \igf{00} \gf{00} + \gf{00} \cdot \sum_{(k,k) \in Steps} z^k
\ee

For $\gf{ll},l=1,\cdots,m-1$, the equation is
\be
\gf{ll} = \gf{00} + \sum_{k=1}^l \igf{l-k,0} \gf{0,l-k}
\ee
\noindent
and for the remainder,
\be
\gf{ij} = \left\{ \begin{array}{cc} 
\displaystyle \sum_{k=1}^i \gf{i-k,0}\igf{k,j} & \textrm{for} \; i < j \\
\displaystyle \sum_{k=1}^j \gf{0,j-k} \igf{i,k} & \textrm{for} \; i \geq j
\end{array} \right.
\ee

For the irreducible generating functions, we need to specify only $\igf{0j}$ and $\igf{j0}$ because
\be
\igf{ij} = \left\{ \begin{array}{cc}
\igf{0,j-i} & \textrm{for} \; i < j \\
\igf{i-j,0} & \textrm{for} \; i \geq j
\end{array} \right.
\ee
\noindent
simply by definition. For $\igf{00}$, we need to consider both starting and ending steps and sum on all possible combinations of these.
\be
\igf{00}  =  \sum_{\substack{(i,j) \\ X[i],Y[j]>0}} & z^{Steps[i]_y +Steps[j]_y} \gf{X[i],Y[j]}
\ee
\ni
where $X[i] = Steps[i]_x-Steps[i]_y-1,Y[j]=Steps[j]_y-Steps[j]_x-1$.

For the remaining irreducible generating functions, we will need to sum over only either on the starting steps or on the ending steps depending on whether we are considering $\igf{0i}$ or $\igf{i0}$.
\be
\igf{0i} & = & \sum_{\substack{k \\ Steps[k]_x-Steps[k]_y>0}} z^{Steps[k]_y} \gf{Steps[k]_x-Steps[k]_y-1,i-1} \\
\igf{i0} & = & \sum_{\substack{k \\ Steps[k]_y-Steps[k]_x>0}} z^{Steps[k]_y} \gf{i-1,Steps[k]_y-Steps[k]_x-1}
\ee

Thus we have, for a given $m$, $2m^2$ algebraic equations in as many variables and we should be able to eliminate everything except $\gf{00}$ and obtain a single polynomial equation in $\gf{00}$. 

\subsection{Infinite Width with Boundary Interaction}
Just as for generating functions of finite width, it is possible to automate the calculation of weight enumerators with one parameter $t$ counting the number of times the walk touches the line $y=x$.

The algorithm here is similar to that of the previous section and in fact, we will need the generating functions $\gf{ij}$ calculated earlier for the same set of steps.

We begin with the same example as in the previous section, namely with steps $\{ [0,1],[1,0],[0,2],[2,0] \}$. Now, let $H(z,t)$ be the weight enumerator. Notice that the only difference between $H$ and $\gf{00}$ is this extra parameter $t$ in the former. We need only one extra equation for $H$ apart from the other ones. This one is almost exactly like the one for $\gf{00}$. Namely,
\be
H = 1 + t \left(z \gf{00} + z^2 \gf{01} + z \gf{10} + z^2 \gf{11} \right) H.
\ee
\ni
All this is saying is that $H$ is either the empty walk or it is composed of all possible starting and ending steps with a factor of $t$ contributing towards the ending step followed by a smaller such walk. The calculations of $\gf{ij}$ have to be done using the same equations as in the previous section.

This gives the following algebraic equation for $H$.
\be
& 1+3 t^2 H^2-2 t z H+3 t H-4 H+t^3 H^3+3 t^3 z^2 H^3+2 t^2 z^2 H^2+6 t z H^2 \cr
&- 3 t^2 z H^2-3 t^3 z^2 H^4+t^4 z^2 H^4+2 t^2 z^2 H^4+2 t^3 z^3 H^4-2 t^4 z^3 H^4 -2 t^3 z^3 H^3\cr
&+ t^4 z^4 H^4+t^4 z H^4-6 H^3 t z-3 t^2 H^4 z-9 t H^2+6 H^2+2 H^4 t z -6 t^2 H^3\cr
&+H^4+9 H^3 t - 4 H^3-t^3 H^4+3 H^4 t^2-3 H^4 t+6 t^2 z H^3-4 t^2 z^2 H^3 =0
\ee
\ni
which is of fourth order in $H$, just as before.

When there is a general set of steps, we have the same equations as before with an additional equation for $H$,
\be
H = 1 + t \cdot \left( \sum_{(k,k) \in Steps} z^k \right) H + t \igf{00} H.
\ee

Solving this system will give us the required algebraic equation for $H$.

\section{Empirical Guessing}
We say that a sequence of rational functions $\{ F_w(z) \}_{w=0}^\infty$ belongs to $\azl{L}(w,z)$ if each function $F_w(z)$ can be written in the form
\be
F_w(z) = \frac{P_w(z)}{P_{w+1}(z)}
\ee
\ni
where the polynomials $\{ P_w(z) \}_{w=0}^\infty$ satisfy a recurrence of order $L$ in $w$ with constant (in $w$) coefficients,
\be
\sum_{i=0}^L a_i(z) P_{w+i}(z) = 0
\ee

For any given set of steps, one can empirically check if the generating functions $\phi_w(z)$ belong to the class $\azl{L}(w,z)$. This is done by using the holonomic ansatz \cite{Z} and searching for a recurrence of order $L$ among the numerators $P_w(z)$.

For most steps, it does turn out that the numerator of $\phi_w(z)$ is precisely the denominator of $\phi_{w-1}(z)$. The generating functions of many classes of steps do turn out to belong to $\azl{L}(w,z)$ for some $L$. For example, steps of the form $\{ [0,1],[n,0] \}$ always lead to generating functions which belong to $\azl{n+1}(w,z)$. Yet another nontrivial example is as follows: Steps of the form $\{ [0,1],[1,0],[n,n+2],[n+2,n] \}$ with $n \geq 0$ belong to $\azl{5}(w,z)$ [AZ]!

It is an open problem whether the generating functions of all set of steps belong to $AZ(w,z)$.

\section{Free Energy}
One can define the free energy for this system as follows. Let $c_w(n)$ denote the number of walks from $(0,0)$ to $(n,n)$. Then the free energy $\kappa_w$ is defined by
\be
\kappa_w = \lim_{n \to \infty} \frac{1}{n} \log c_w(n)
\ee

In general, this is the smallest real positive singularity of the generating function. Since the generating function for the walk with any fixed $w$ is rational, one can expand it in partial fractions and this number is given by the natural logarithm of the largest zero of the denominators.

There are well-established algorithms \cite{SZ,C}, to find the recurrence relation satisfied by the sequence $\{ a_n \}$ given the generating function $\phi(z) = \sum a_n z^n$. We denote this operator as $P(N,n)$. In Maple, this is implemented in the package titled GFUN \cite{SZ}. Now, given this polynomial, there exists an algorithm \cite{BT,WiZ} to find the asymptotic behaviour of the generating function. This is implemented in the package GuessHolo2 \cite{Z}.

One can therefore calculate the free energy for walks with a given set of steps at any finite width as well as for infinite width.

One can also calculate the free energy as a function of the variables $t,s$. Unfortunately there is no explicit formula for the free energy as a function of $t,s$, but given any specific values of $t$ and $s$, these can be calculated exactly. This is mainly because we do not have a general formula for the roots of polynomials of degree $ \geq 5$. For the same reason, one cannot calculate the free energy for the infinite width case as a function of the parameter $t$.

\section{Force}
The force exerted on the plates is given, in the discrete case, by
\be
F(w) = \kappa_{w+1}- \kappa_w
\ee

In the limit when $w$ is very large, the force is defined by the derivative of the free energy with respect to $w$. One can also calculate the force for any specific values of the variables $t,s$. For the same reason as in the previous section, this calculation cannot be done keeping $t$ and $s$ arbitrary.

The structure of the phase diagram is most clearly seen by plotting the force as a function of $t,s$. The region where the force is positive is the desorbing region and where the force is negative is the adsorbing region.

\section{Examples}
We use the algorithms outlined above and some others to calculate quantities of interest for different sets of steps. We emphasize that these are not individually calculated for these particular set of steps but are simply outputs of algorithms described in previous sections. These algorithms are implemented in Maple.

\subsection{$\{ (1,1),(1,-1) \}$ Steps}
We repeat some of the calculations in \cite{BORW} with standard Dyck steps to demonstrate the power of this approach. The corresponding ballot steps are $[0,1],[1,0]$.

The generating function at any finite width can be calculated for even reasonably large widths in short enough times. For example, when $w=3$,
\be
\phi_3(z) = \frac{1-2z}{1-3z+z^2}
\ee

We can calculate the equation satisfied by infinite the width generating function.
\be
-F(z)+1+z F(z)^2 = 0
\ee

Let $P_w(z), Q_w(z)$ be the numerators and denominators of $\phi_w(z)$. One can conjecture a recurrence relation in $w$ for the $Q_w$'s. For these steps, it turns out that
\be
zQ_w(z) - Q_{w+1}(z) + Q_{w+2}(z) = 0
\ee

\begin{figure}[!p]
\begin{center}
\includegraphics[height=10cm]{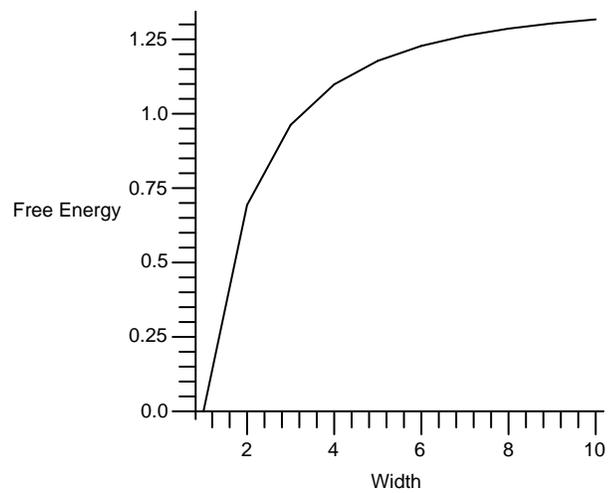}
\caption{Free energy for steps $\{ [0,1],[1,0] \}$ plotted from $w=1$ to $w=10$}
\label{dyckfe}
\end{center}
\end{figure}

\begin{figure}[p]
\begin{center}
\includegraphics[height=10cm]{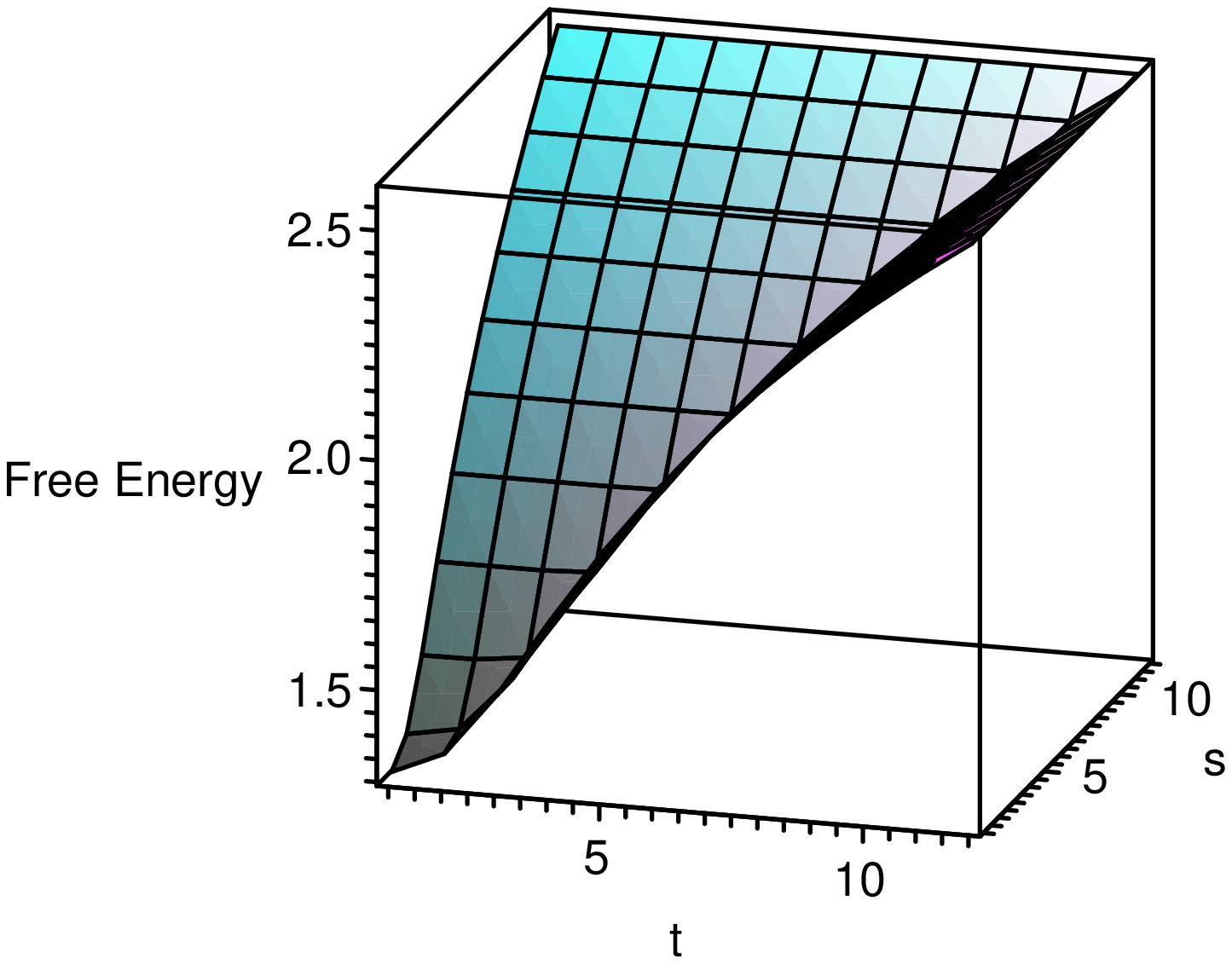}
\caption{Free energy for steps $\{ [0,1],[1,0] \},w=10$ and $t,s=1,\cdots,12$}
\label{dyckfewe}
\end{center}
\end{figure}

Figure \ref{dyckfe} shows the free energy plotted for a range of widths.
 We also plot the free energy for a particular width for a set of weight-enumerating parameters in Figure \ref{dyckfewe}. Notice that even such a small value of $w$ shows the characteristics of the phase diagram in Figure 7 of \cite{BORW}. For $0 \leq t,s \leq 2$, the value of the free energy is more or less constant and outside, it seems to grow more or less linearly and we can clearly notice the non-analyticity at the line $t=s$.

\begin{figure}[p]
\begin{center}
\includegraphics[height=10cm]{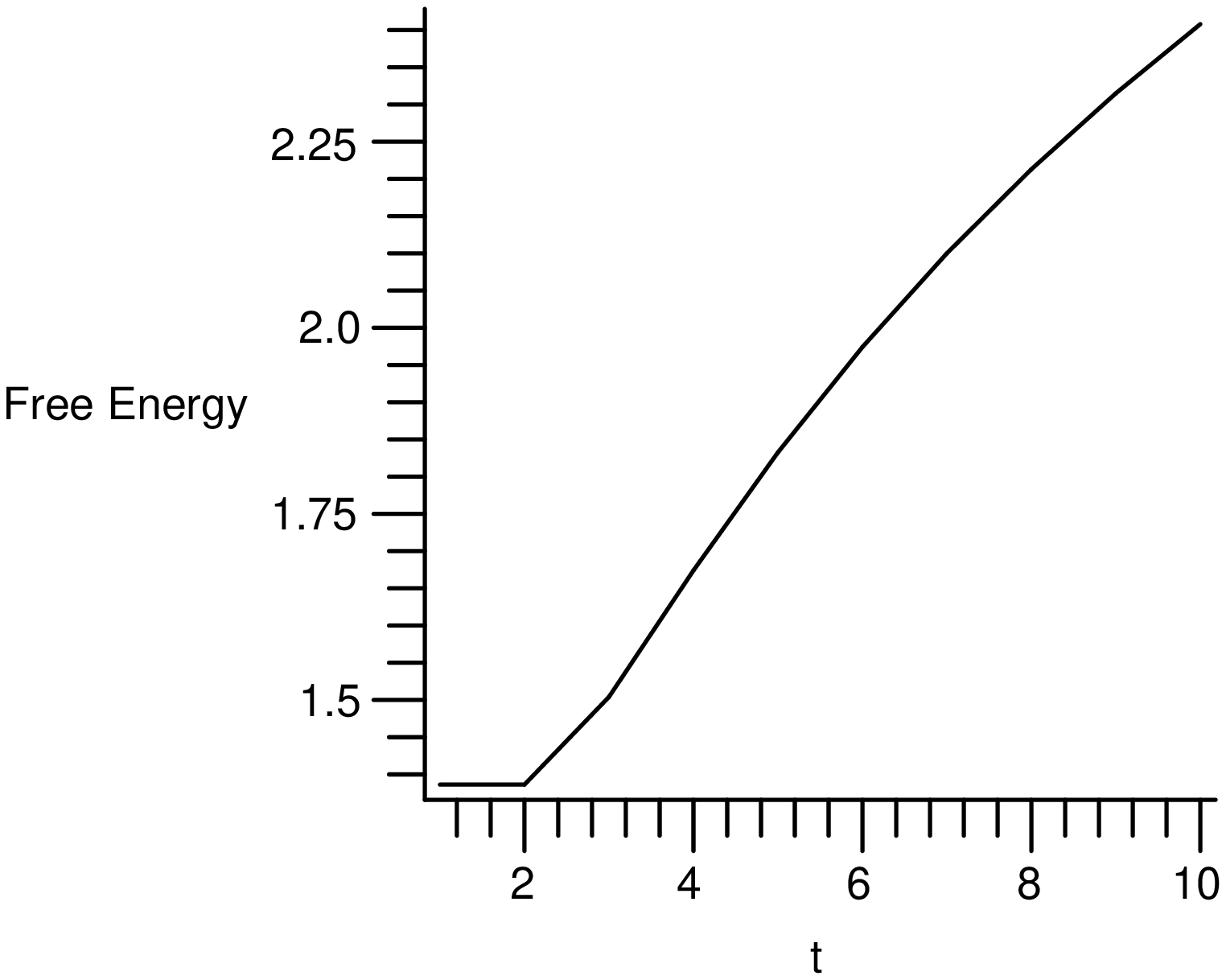}
\caption{The infinite width free energy for steps $\{ [0,1],[1,0] \}$, and $t=1,\cdots,10$}
\label{dyckinffe}
\end{center}
\end{figure}

\begin{figure}[p]
\begin{center}
\includegraphics[height=10cm]{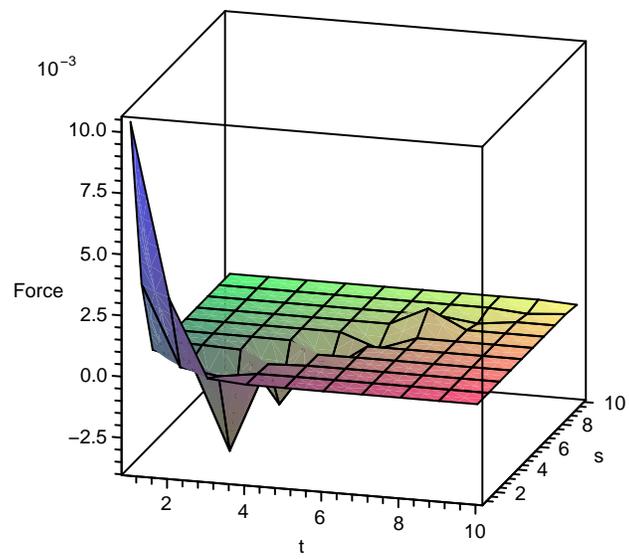}
\caption{Force for steps $\{ [0,1],[1,0] \},w=10$ and $t,s=1,\cdots,10$}
\label{dyckforce}
\end{center}
\end{figure}

We can also plot the free energy of the infinite width case as a function of the parameter $t$ which counts the number of times the walk touches the diagonal $x=y$. This is done in Figure \ref{dyckinffe}.
 To see how the phase diagram looks, one can also plot the force for a reasonable large value of the width. As seen in Figure \ref{dyckforce}, this looks like the derivative of Figure \ref{dyckfewe}. There is a strong positive force in the region $0 \leq t,s \leq 2$, which is the desorbed region. There is also the clear attraction regime around the line $t=s$ as shown in Figure 8 of \cite{BORW}.

\newpage

\subsection{$\{ (1,1),(2,2),(1,-1),(2,-2) \}$ Steps}
\begin{figure}[p]
\begin{center}
\includegraphics[height=10cm]{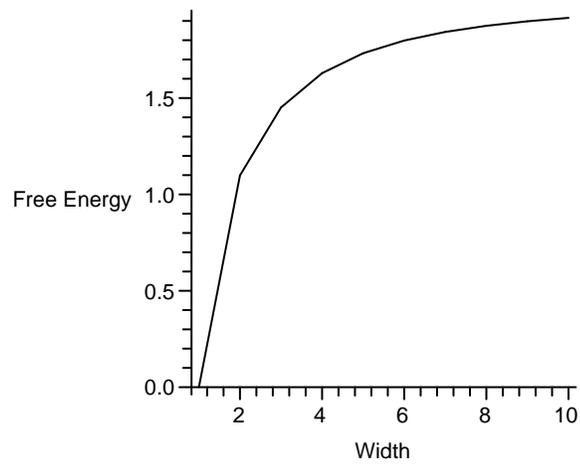}
\caption{Free energy for steps $\{ [0,1],[1,0],[2,0],[0,2] \}$ plotted from $w=1$ to $w=10$}
\label{bballfe}
\end{center}
\end{figure}

For a more nontrivial example, consider the following ballot steps: $\{ [1,0]$, $[2,0], [0,1],[0,2] \}$ \cite{AZ}. Figure \ref{bballfe} shows the free energy plotted for a range of widths. Notice that the behaviour is very similar to that of the previous example in Figure \ref{dyckfe}. We know that the behavior must be monotonically increasing with the width and is a concave function, which is true for both figures. 

\begin{figure}[p]
\begin{center}
\includegraphics[height=10cm]{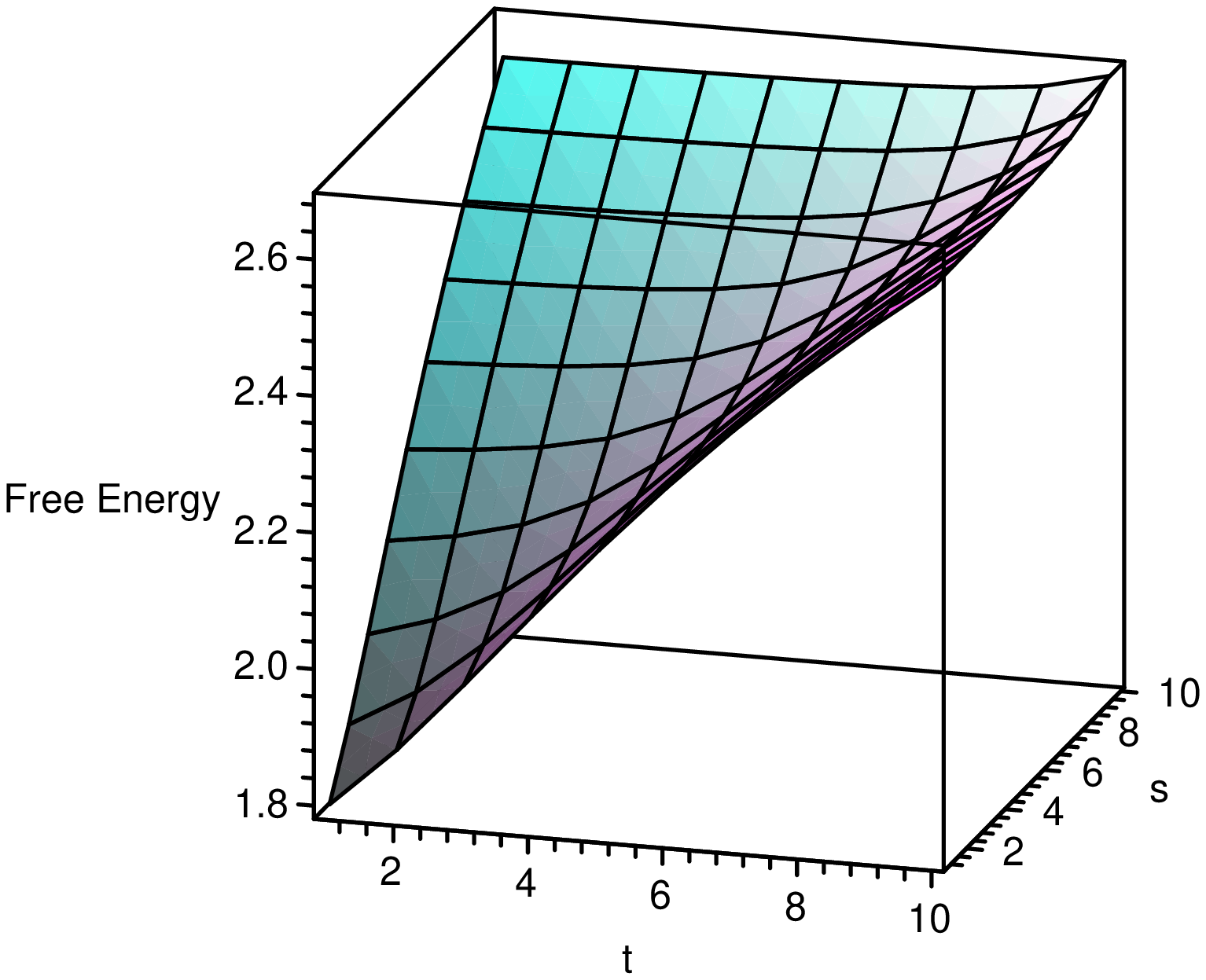}
\caption{Free energy for steps $\{ [0,1],[1,0],[2,0],[0,2] \},w=6$ and $t,s=1,\cdots,10$}
\label{bballfewe}
\end{center}
\end{figure}

We also plot the free energy for a particular width for a set of weight-enumerating parameters in Figure \ref{bballfewe}. Here too, the free energy increases with increasing values of $t,s$. However, unlike Figure \ref{dyckfewe}, there is no apparent loss of analyticity. That might be because the width is too small here. 

\begin{figure}[p]
\begin{center}
\includegraphics[height=10cm]{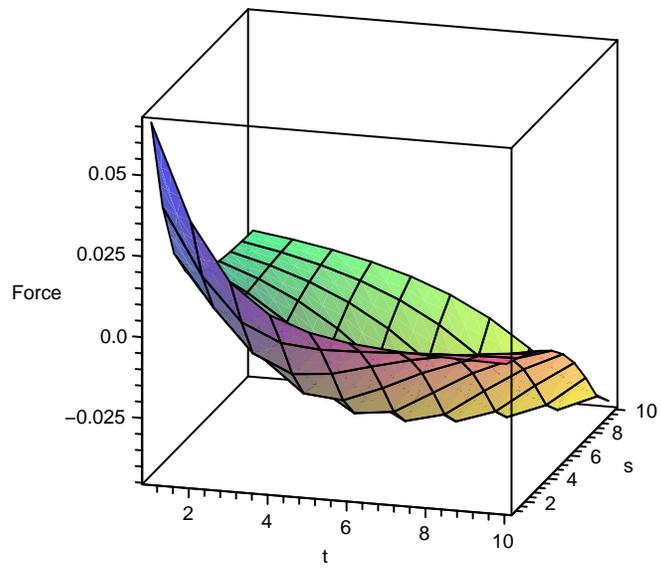}
\caption{Force for steps $\{ [0,1],[1,0],[2,0],[0,2] \},w=5$ and $t,s=1,\cdots,10$}
\label{bballforce}
\end{center}
\end{figure}

To see how the phase diagram looks, one can also plot the force for a reasonable large value of the width. Figure \ref{bballforce} shows this. The sheet in this figure is smoother than in the analogous sheet in the previous example Figure \ref{dyckforce}. And there is again a similarity between the two figures simply because there is a similarity between the corresponding plots for the free energy.

\newpage

\section{Acknowledgements}
The authors would like to thank Jacques Carette for help in programming with Maple, and Joel Lebowitz and Mireille Bousquet-Melou for discussions. The work of the first author was supported in part by NSF grant DMR-044-2066 and AFOSR grant AF-FA 9550-04-4-22910.

\appendix

\section{Usage}
Here we describe the basic procedure for using the program to determine the necessary information for your favorite polymer.

First off, download the package \emph{POLYMER} from the webpage of Doron Zeilberger or by downloading the source of this paper from \\ \texttt{http://arXiv.org/cond-mat/0701674} . Start Maple and at the prompt, type 
\bdm
> \quad \mathrm{read \; `POLYMER`:}
\edm

If you start Maple in a different directory, you have to specify the path where you saved the package. For example, if you saved it in \texttt{C:$\setminus$Packages} or in \texttt{/tmp/Packages} (depending on the OS), type
\be
& > & \quad \mathrm{read \; `C:\setminus\setminus Packages \setminus\setminus POLYMER`:} \nonumber \\
& > & \quad \mathrm{read \; `/tmp/Packages/POLYMER`:} \nonumber
\ee

To see the list of programs, type
\bdm
> \quad \mathrm{Help();}
\edm

We now describe the main tools of the package. The basic syntax is as follows. Any point $(x_1,y_1)$ is represented as $[x_1,y_1]$. The set of steps is represented within curly braces. For example, the steps shown in Figure \ref{walk0to2} are depicted by $\{ [0,1],[1,0] \}$.

\subsection{Walks}
The most basic program in the package is the one that computes the number of walks from any point $(x_1,y_1)$ to any other point $(x_2,y_2)$ using any set of steps and any width $w$. 

\subsubsection{Simple Walks}
For example, to see the number of ways of getting from the origin to the point $(2,2)$ using the steps above with the constraint given by $0 \leq x-y \leq 3$ is
\be
> &\quad \mathrm{polymerBE(\{[0,1],[1,0]\},[0,0],[2,2],3);}& \nonumber \\
&2& \nonumber
\ee
while the same walk with the stronger constraint $0 \leq x-y \leq 1$ is 
\be
> &\quad \mathrm{polymerBE(\{[0,1],[1,0]\},[0,0],[2,2],1);}& \nonumber \\
&1& \nonumber
\ee
To see why that is true, look at Figure \ref{walk0to2}.

\subsubsection{Walks with Boundary Interactions}
We repeat the calculation for exactly the same situation in the cases where the width, $w=1,2,3$.
\be
> &\quad \mathrm{WEpolymerBE(\{[0,1],[1,0]\},[0,0],[2,2],1,t,s);}& \nonumber \\
&t^2 s^2& \nonumber
\ee
\noindent
This is because the only walk touches both walls twice. For $w=2$,
\be
> &\quad \mathrm{WEpolymerBE(\{[0,1],[1,0]\},[0,0],[2,2],2,t,s);}& \nonumber \\
&t^2+ts& \nonumber
\ee
which is because there are two walks now and one walk does not touch the wall on the far right at all. And for $w=3$,
\be
> &\quad \mathrm{WEpolymerBE(\{[0,1],[1,0]\},[0,0],[2,2],3,t,s);}& \nonumber \\
&t^2+t& \nonumber
\ee
which is because neither of the two walks touches the wall on the far right.

\subsection{Generating Functions}
The package can be used to compute this generating function for any finite width as well as the special case of the infinite width. 

As a simple example, consider the same steps as before. Then, for $w=1$, there is only one way of getting to the point $(n,n)$, which is by the zigzag route extending the walk on the left of Figure \ref{walk0to2}. Therefore, the generating function is given by 
\be
\phi_1(z) & = & 1+z+z^2+z^3+\cdots \cr
          & = & \frac{1}{1-z}
\ee

To verify this, type
\be
> &\quad \mathrm{rigorgf(\{[0,1],[1,0]\},1,z);}& \nonumber \\
&\displaystyle \frac{1}{1-z}& \nonumber
\ee

For a more nontrivial example, see what happens for $w=3$.
\be
> &\quad \mathrm{rigorgf(\{[0,1],[1,0]\},3,z);}& \nonumber \\
&\displaystyle \frac{1-2z}{1-3z+z^2}& \nonumber
\ee
To get the number of walks up to $(n,n)$, one simply needs to look at the $n$th Taylor coefficient.

We can also calculate the generating functions of walks with variables $t,s$. As an example, we take the same steps as before with $w=1$
\be
> &\quad \mathrm{rigorgfWE(\{[0,1],[1,0]\},3,z,t,s);}& \nonumber \\
&\displaystyle \frac{1-z-sz}{1-z-sz-tz+stz^2}& \nonumber
\ee

For the case of infinite width, one can again calculate the generating function. The program returns the polynomial equation that it satisfies.
\be
> &\quad \mathrm{RGF2D(\{[0,1],[1,0]\},z,F);}& \nonumber \\
&\displaystyle \{ 1-F+z F^2 \}& \nonumber
\ee

This means that $F(z)$ satisfies the equation $1-F(z)+zF^2(z)=0$. Since this is a quadratic equation in $F$, it can be solved easily.
\be
F(z) = \frac{1 \pm \sqrt{1-4z}}{2z}
\ee

Since we want a formal power series and the taylor coefficients to be non-negative, we take the negative root. Taking the Taylor expansion gives
\be
&> \quad \mathrm{taylor((1-sqrt(1-4 z))/(2 z),z=0,10);}& \nonumber \\
&\displaystyle 1+ z+2z^2+5z^3+14z^4+42z^5+132z^6+429z^7+1430z^8+O(z^9) & \nonumber
\ee

These coefficients are precisely the \emph{Catalan numbers} [S]. 

One can also calculate the weight enumerator for the same set of walks with infinite width, where $t$ is the parameter whose coefficient counts the number of times the walk touches the diagonal.
\be
> &\quad \mathrm{RGF2DWE(\{[0,1],[1,0]\},z,F,t);}& \nonumber \\
&\displaystyle \{1+(t-2)F+(t^2z+1-t)F^2  \}& \nonumber
\ee
\ni
and plugging in $t=1$ gives the unweighted generating function, as expected.

\subsection{Empirical Guessing}
That is, let $P(w,W)$ be an operator where $W$ acts by shifts: $W Q_w(z) = Q_{w+1}(z)$. The degree of $W$ in $P$ is called the \emph{order of the recurrence} and the degree of $w$ in $P$ is called the \emph{degree of the recurrence}. Note that $P$ implicitly depends on $z$.

Suppose we have a walk with steps $[0,1],[1,0]$. Let us try to find a recurrence of order 2 and degree 0 as [BORW] suggests.
\be
&> \quad \mathrm{stepsrec(\{[1,0],[0,1]\},0,2,z,w,W);}& \nonumber \\
&\displaystyle z-W+W^2 & \nonumber
\ee

Similarly, we can find recurrences for the weight enumerators. The previous steps satisfy exactly the same recurrence for their weight enumerators! Consider the walk with steps $[0,1],[1,1],[1,0]$ and order 4 and degree 0, we find
\be
&> \quad \mathrm{stepsrecWE(\{[1,0],[1,1],[0,1]\},0,2,z,w,W,t,s);}& \nonumber \\
&\displaystyle W^2+(z-1)W+z & \nonumber
\ee

We remind the reader that these are essentially empirical results. To actually prove these, one has to write down the corresponding nonlinear recurrence relations for the generating functions $\phi_w(z)$ and prove them on a case-by-case basis.

\subsection{Free Energy}
The package can be used to calculate the free energy for any specific width, plot the free energy (using the weight enumerating generating function) at a specific width for ranges of $t,s$ as well as plot the ordinary generating function in a range of widths.

Suppose we want to calculate the free energy for a specific width. As an example, consider the same steps and $w=3$.
\be
> &\quad \mathrm{FE(\{[0,1],[1,0]\},3);}& \nonumber \\
&\displaystyle ln \left( \frac{3}{2} + \frac{\sqrt{5}}{2} \right)& \nonumber
\ee

One could also, for example, plot the free energies for the same walk from widths of 1 to 10.
\be
> &\quad \mathrm{plotFE(\{[0,1],[1,0]\},1,10);}& \nonumber \\
& \displaystyle \mathrm{The \; asymptotic \; value \; is \;} 1.386294361 & \nonumber
\ee

The output is Figure \ref{dyckfe}.

For the weight enumerators, one can plot free energies for a fixed width and range of $t$ and $s$ parameters. Unfortunately, we cannot get asymptotic values here. For instance, with the same steps as before, we can look at the case when $w=3$ and the range $t=1,...,10, s=1,...,10$. The
\be
> &\quad \mathrm{plotFEWE(\{[0,1],[1,0]\},10,1,12,1,12);} \nonumber
\ee

The output is Figure \ref{dyckfewe}.

And lastly, one can plot the free energy as a function of the variable $t$ for the infinite width case.

\be
> &\quad \mathrm{plotinfFE(\{[0,1],[1,0]\},1,10);} \nonumber
\ee

The output is Figure \ref{dyckinffe}.

\subsection{Force on Walls} 
Using essentially the same algorithm as for the free energy, one can plot the force to get an idea of the adsorption/desorption phase diagram.

For example, for the walk with steps $(1,0),(0,1)$ and $w=10$, we can plot the force in the range $t,s=1,\cdots,10$.
\be
> &\quad \mathrm{ForceWE(\{[0,1],[1,0]\},10,1,10,1,10);} \nonumber
\ee

The output is Figure \ref{dyckforce}.


\begin{thebibliography}{99}

\bibitem[AZ]{AZ} Arvind Ayyer and Doron Zeilberger, {\it The Number of [Old-Time] Basketball games with Final Score n:n where the Home Team was never losing but also never ahead by more than w Points}, Elec. J. of Combinatorics, {\bf 14(1)} (2007), R19.
\bibitem[Bo]{Bo} Mireille Bousquet-Melou, {\it Discrete Excursions}, Preprint located at \texttt{http://www.arXiv.org/abs/math.CO/0701171}
\bibitem[BORW]{BORW} R. Brak, A.L. Owczarek, A. Rechnitzer, S.G. Whittington, {\it A directed walk model of a long chain polymer in a slit with attractive walls}, J. Phys. A, {\bf 38}, 2005, 4309-4325.
\bibitem[BT]{BT} G. Birkhoff, {\it Formal theory of irregular difference equations}, Acta Math., {\bf 54}, 1930, 205-46,\\
G. Birkhoff,  W. Trjitzinsky, {\it Analytic theory of singular difference equations}, Acta Math., {\bf 60}, 1932, 1-89.
\bibitem[C]{C} L. Comtet, {\it Calcul pratique des coefficients de Taylor d'unefonction algebrique}. L'Enseignement Mathematique {\bf 10}, 1964, 267-70.
\bibitem[Du]{Du} Philippe Duchon, {\it On the enumeration and generation of generalized Dyck words. Formal power series and algebraic combinatorics},  Discrete Math.  {\bf 225}  (2000),  no. 1-3, 121-135.
\bibitem[R]{R} E.J. Janse van Rensburg, {\it  The statistical mechanics of interacting walks, polygons, animals and vesicles}, Oxford Lecture Series in Mathematics and its Applications, 18. Oxford University Press, Oxford, 2000.
\bibitem[SZ]{SZ} Bruno Salvy, Paul Zimmerman, {\it GFUN: a Maple package for the manipulation of generating and holonomic functions in one variable}, ACM Transactions on Mathematical Software, {\bf 20,2}, 1994, 163-77.
\bibitem[WiZ]{WiZ} Jet Wimp and Doron Zeilberger, {\it Resurrecting the asymptotics of linear recurrences}, J. Math. Anal. Appl., {\bf 111}, 1985, 162-76.
\bibitem[Z]{Z} Doron Zeilberger, {\it The HOLONOMIC ANSATZ I. Foundations and Applications to Lattice Path Counting}, submitted. Preprint at \\
http://www.math.rutgers.edu/$\sim$zeilberg/ \\
mamarim/mamarimhtml/ansatzI.html.

\end{thebibliography}
\end{document}